\documentclass{emulateapj}

\shorttitle{Expected number of massive galaxy relics in the present-day Universe}
\shortauthors{V. Quilis \& I. Trujillo} 

\def\gsim{ \lower .75ex \hbox{$\sim$} \llap{\raise .27ex \hbox{$>$}} }
\def\lsim{ \lower .75ex \hbox{$\sim$} \llap{\raise .27ex \hbox{$<$}} }

\begin{document}
\title{Expected number of massive galaxy relics in the present-day Universe}

\author{Vicent Quilis} 
\affil{Departament d'Astronomia i Astrof\'{\i}sica, 
Universitat de Val\`encia, 46100 - Burjassot, Val\`encia, Spain}
\email{vicent.quilis@uv.es}
\and
\author{Ignacio Trujillo\altaffilmark{1}} 
\affil{Instituto de Astrof\'{\i}sica de Canarias, 
c/ V\'{\i}a L\'actea s/n, E38205 - La Laguna, Tenerife, Spain}
\altaffiltext{1}{Departamento de Astrof\'isica,
Universidad de La Laguna, E-38205 La Laguna, Tenerife, Spain}

\begin{abstract}

The number of present-day massive galaxies that has survived untouched since their formation at high-z is an
important observational constraint to the hierarchical galaxy formation models. Using three different semianalytical
models based on the Millenium simulation, we quantify the expected fraction and number densities of the massive
galaxies form at z$>$2 which have evolved in stellar mass less than 10\% and 30\%. We find that only a small
fraction of the massive galaxies already form at z$\sim$2 have remained almost unaltered since their formation
($<$2\% with $\Delta$M$_\star$/M$_\star$$<$0.1 and $<$8\% with $\Delta$M$_\star$/M$_\star$$<$0.3 ). These fractions
correspond to the following  number densities of massive relics in the present-day Universe: $\sim$1.2 $\times$
10$^{-6}$ Mpc$^{-3}$ for  $\Delta$M$_\star$/M$_\star$$<$0.1 and $\sim$5.7$\times$10$^{-6}$ Mpc$^{-3}$ for 
$\Delta$M$_\star$/M$_\star$$<$0.3.  The observed number of relic candidates found in the nearby Universe is today
pretty uncertain (with uncertainties up to a factor of $\sim$100) preventing to establish a firm conclusion about
the goodness of current theoretical expectations to predict such important number.

\end{abstract}

\keywords{dark matter --- galaxies: halos --- galaxies: formation 
--- galaxies: evolution}

\section{Introduction}

Merging of galaxies is an intrinsically stochastic process. That means that there should be a number of galaxies
formed in the early epochs of the Universe that  has remained  untouched until today. Identifying and exploring
these relics objects is of fundamental relevance to understand the conditions and properties of the primordial
phases of galaxy formation. But, how many of these objects are theoretically expected to survive without being
significantly modified since high redshift? In this letter, we address this issue focusing our attention to the most
massive (M$_\star$$>$8$\times$10$^{10}$\,M$_{\odot}$)   galaxies in the Universe.

Observationally, massive galaxies have been found to be more compact at high redshift  (Daddi et al. 2005; Trujillo
et al. 2006). Particularly those with spheroid-like morphologies show sizes (measured using their effective radii) a
factor of $\sim$4 more smaller at z$\sim$2 (see e.g. Trujillo et al. 2007; Buitrago et al. 2008). These massive
compact galaxies are expected to grow as cosmic time increases both in stellar mass and in size by a continuous
accretion of minor satellites (e.g.  Bezanson et al. 2009; Hopkins et al. 2009). This channel of evolution agrees
very well with many observations and it has been proven theoretically able to produce the expected size evolution
(e.g. Naab et al. 2009; Sommer-Larsen \& Toft 2010, Feldmann et al. 2010, Oser et al. 2012). Being the merging a
stochastic process, it is expected that a number of these massive compact galaxies remain unaltered since their
formation. Consequently, quantifying the number of these massive old compact objects in the present-day Universe is
a  proxy to explore the number of massive relics  of the early Universe.

Contradictory claims in relation to the number of today massive compact galaxies with old stellar populations have
been reported. On one hand, Trujillo et al. (2009) using the Sloan Digital Sky Survey (SDSS) did not find any clear
evidence for a single massive "old" galaxy which has survived untouched since their high-z formation. In fact, they
found that the number of massive (M$_\star$$>$8$\times$10$^{10}$\,M$_{\odot}$) and compact (r$_e$$<$1.5 kpc)
galaxies in the nearby Universe (z$<$0.2) is less than 0.03\% (see also Taylor et al. 2010 for a confirmation of
this result). Moreover, these galaxies are not old but relatively young  ($\sim$2 Gyr; see also Ferr\'e-Mateu et al.
2012). On the other hand, Poggianti et al. (2013) using a much more modest area find evidence for up to 4 old massive
compact galaxies satisfying the above selection criteria of stellar mass and size. The reason behind these two
conflicting results remains a mystery and further studies are necessary to clarify this point.

From the theoretical perspective, there are not single quantification about the expected number of massive galaxies
which should not have suffered any grow since their formation. Answering quantitatively this question is crucial  if
we want to settle the minor merging scenario (the present-day favoured channel of growth)  under solid grounds.
Intuitively, a moderate low number of massive compact galaxies in the nearby Universe should favour the hypothesis
that massive galaxies have evolved by satellite accretion.  However, if the number of today relics is very small or
nil this will put into question the minor merging hypothesis. This is again due to the intrinsic stochastic nature
of the merging model. It is then an urgent question to quantify which is the exact theoretical prediction about the
expected number of massive relics in the present-day universe. Fortunately, current cosmological simulations are
large enough to permit the estimation of this fraction with accuracy.

In this Letter, we explore the model predictions on the fraction and number density of massive galaxies form at
z$>$2 which have not increased their stellar mass since that epoch by more than 10\% and 30\%. To quantify these
numbers we have used  three different semi-analytical models (DeLucia et al. 2007, Guo et al. 2011 and Guo et al.
2013) based on the Millennium simulation.

\section{Galaxy catalogues}

We use the public release of  a two very large N-body
simulations, Millennium I~\citep{springel05} (MI) and Millennium I-WMAP7~\citep{guo13} (MI7), a version of the original Millenium I simulation run using the seven-year WMAP data \citep{komatsu11}. The cosmological parameters for MI (MI7) simulation are: $\Omega_m=0.25$ ($\Omega_m=0.272$), $\Omega_b=0.045$($\Omega_b=0.045$), $\Omega_{\Lambda}=0.75$($\Omega_{\Lambda}=0.728$),
$n=1$($n=0.961$), $\sigma_8=0.9$($\sigma_8=0.807$), $H_0=73 \,km s^{-1}Mpc^{-1}$($H_0=70.4 \,km s^{-1}Mpc^{-1}$). The two simulations use
the same number of particles, $2160^3$. Thus, the computational box has sides of
$685\,Mpc$ ($710\,Mpc$ ), and particles masses of $1.18\times10^9\,M_{\odot}$ ($1.32\times10^9\,M_{\odot}$). In order to facilitate 
the comparison between the three different catalogues and the observational data (see Sec. 4), we present all the results assuming 
$H_0=70 \,km s^{-1}Mpc^{-1}$.


A combination of two halo finders, a friends-of-friends (FoF) by ~\cite{davis85} and SUBFIND~\citep{springel01}, is used 
in order to analyze the simulations and to build up the dark matter merger trees.   The dark matter haloes found in the N-body
simulations are transformed into galaxies according to different semi-analytical models. Depending on the particular 
implementation of each model, several phenomenological recipes are used to produce the gas and stellar components in 
the virtual galaxies. In this Letter, we use three semi-analytical models available in the  Millennium database web ~\citep{lemson06}.

The first one is the model by \cite{delucia07} and the other two, in fact they are the same model but applied to different simulations, are \cite{guo11} and \cite{guo13}. All of them  are very similar
being the models by \cite{guo11} and \cite{guo13} an improvement of the model by \cite{delucia07}.
The Guo's models implement several new features:
the separate evolution of sizes and orientations of gaseous and stellar discs, the
size evolution of spheroids, tidal and ram-pressure stripping of satellite galaxies, and the disruption
of galaxies to produce intra-cluster light. 
All three models include the effects of AGN feedback. 
The stellar masses of the semi-analytical galaxies were estimated assuming a Chabrier (2003) initial mass function (IMF) for the three
models. 
This is consistent with the IMF assumed in the observational works used to compare with the 
theoretical predictions.

\begin{figure}
\includegraphics[scale=.5]{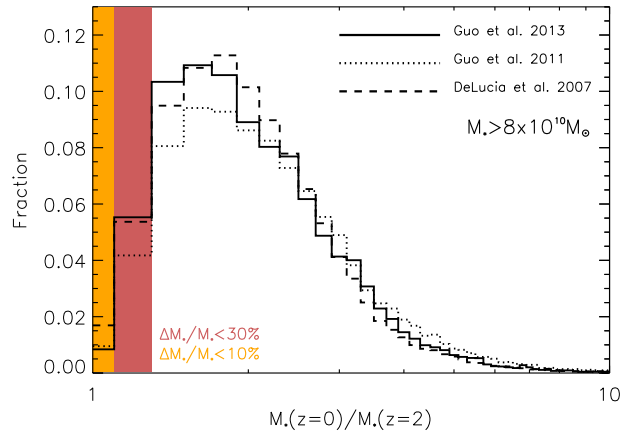}
\caption{Fraction of present-day massive galaxies that already existed at z$\sim2$ as function of their relative increase of stellar mass.The different line styles stand for the three considered models. The color shaded areas  
show the fraction of galaxies that have increased their stellar masses less than a 10\% and 30\%, respectively.}
\label{fig1}
\end{figure}

 \section{Results}

We generate three galaxies catalogues using the 
Millennium database web ~\citep{lemson06}. Each catalogue corresponds to one different semi-analytical
model as previously discussed. For each of the catalogues, we select all the 
massive galaxies at present-day as those objects with stellar masses
between $8\times 10^{10}\,M_{\odot}$ 
and $10^{13}\,M_{\odot}$.

We define a relic as a galaxy in the present-day Universe that has barely increased its stellar mass since $z\sim 2$. In order to identify the possible candidates to relic galaxies, we use the merger tree structures to trace backwards in time the massive galaxies
identified at $z\sim 0$, together with two conditions: i) the galaxy must be already form at $z\sim2$, and ii) the galaxy mass at $z\sim 2$ has to be larger than the 90\% or the 70\%, depending on the considered case, of the limit mass at the present-day $8\times 10^{10}\,M_{\odot}$. Once all the candidate to relics are selected, it is possible to obtain their stellar mass increment at several redshift since 
$z\sim2$.   
To have an estimation of the number of massive relics in the present-day Universe, we quantify the number of almost pure massive relics
 as the number of galaxies, among all the candidates, with an increase in  their stellar mass less than 10\% since z=2. Additionally, we also explore the number of massive galaxies in the z=0 Universe with an increase in stellar mass less than 30\% since z=2.  This last limit is selected to avoid that the galaxy has suffered a major (i.e. 1:3 or lower ratio) merger since z=2 that could have altered their structure. 

In Figure~\ref{fig1}, we show the fraction of massive galaxies  at $z\sim 0$ already present at $z\sim2$  as a
function of the relative increment of their stellar mass. All three considered models, show that roughly half  of
the massive galaxies already build at $z\sim2$ have increased their masses at $z\sim0$ more than  a factor of two. 
Only a small fraction of the massive galaxies at the present epoch corresponds to massive galaxies at $z\sim2$ that 
remain untouched or with a minor increase in their masses. In fact, less than a 2\% (8\%) of the massive
galaxies already form at z$\sim$2 gained less than a 10\% (30\%) of their stellar
mass since that epoch.

\begin{figure*}
\includegraphics[scale=.9]{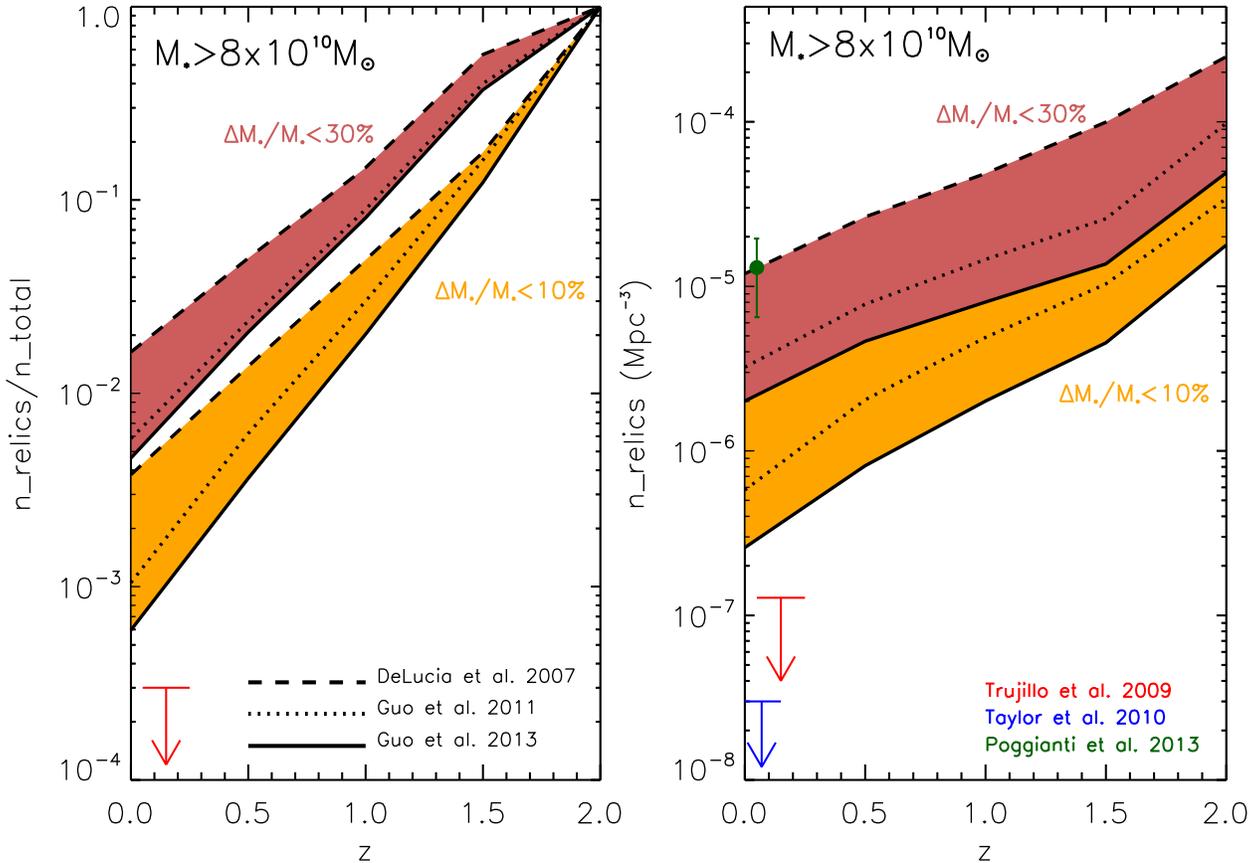}
\vspace{1 cm}
\caption{Left panel: redshift evolution of the ratio of the relic galaxies to the total number of massive galaxies. The three different lines represent the three considered models. 
Coloured areas,  orange (orange-red)  show galaxies that have increased their masses less
than a 10\% (30\%) since $z\sim2$. Right panel: redshift evolution of the comoving number density of relic galaxies. The lines and colour shaded areas stand as for the right panel.  The red and blue arrow show the observational upper limits from Trujillo et al. (2009) and Taylor et al. (2010), respectively. The green point display the observational data from Poggianti et al. (2013).} 
\label{fig2}
\end{figure*}

\begin{table*}
 \caption{Fractions and number densities of the present-day massive galaxy relics}
  \label{tab:distance}
  \begin{center}
   \leavevmode
    \begin{tabular}{lllll} \hline \hline              
  $\Delta M_*/M_*$   &  DeLucia et al. (2007)  &   Guo et al. (2011)  &  Guo et al. (2013) & Average  \\ 
     (since z=2) &   &   &  & \\\hline
&$\rm{n_{relics}}/n_{total} $ at $z\sim 0$& & & \\ \hline
$<$10\%& 0.0038 & 0.001 & 0.0006 & 0.0018 \\
$<$30\%& 0.016 & 0.006 & 0.005 & 0.009 \\ \hline
& Number density  (Mpc$^{-3}$)& & &\\ \hline
$<$10\%& 2.7$\times$10$^{-6}$& 5.8$\times$10$^{-7}$&  2.6$\times$10$^{-7}$ & 1.2$\times$10$^{-6}$ \\
$<$30\%&  1.2$\times$10$^{-5}$& 3.2$\times$10$^{-6}$ &  2.0$\times$10$^{-6}$ & 5.7$\times$10$^{-6}$ \\ \hline
  
    \end{tabular}
  \end{center}
\end{table*}

In Figure~\ref{fig2}, we represent the redshift evolution of the ratio of the number of relic galaxies to the total
number of massive galaxies (left panel) and the number density of relics for the three considered models (right
panel). The total number of massive galaxies at each redshift account for all the massive galaxies that are
progressively being incorporated to the family of massive galaxies as cosmic time increases. No substantial
differences appear between the three models. At all redshifts, \cite{delucia07} model produces ratios of relic
galaxies slightly higher than the other  two models. As in Fig~\ref{fig1}, we distinguish between two samples, those
massive galaxies that have increased their relative stellar  masses less than a 10\% and those others with a
relative mass increment less than a 30\%. For the shake of completeness, we include the  observational upper limits
from Trujillo et al. (2009) and Taylor et al. (2010), and the observational data from Poggianti et al. (2013)
\footnote{The number densities from Trujillo et al. (2009), Taylor et al. (2010) and Poggianti et al. (2013) are
computed assuming a standar cosmology: $\Omega_m=0.3$, $\Omega_{\Lambda}=0.7$, and $H_0=70 \,km s^{-1}Mpc^{-1}$.}.

The nowadays ratios of relic galaxies to total number of galaxies and the present-day number densities  for the
three different considered catalogues are summarized in Table 1.

\section{Discussion}

As mention in the Introduction, to have an accurate estimation of the expected number density of massive relics in
the present-day Universe is crucial to probe the minor merging channel of galaxy growth. Being the massive galaxies
at high-z much more compact compared to current massive counterparts, a good proxy to identify massive galaxy relics
in the nearby Universe is to search for massive compact galaxies with old stellar populations. The observed fraction
of compact (r$_e$$<$1.5 kpc) massive (M$_\star$$>$8$\times$10$^{10}$\,M$_{\odot}$) galaxies with z$<$0.2 in the
present-day Universe (i.e. a good set of candidates to be relics of the early Universe) is $\sim$0.0003 (Trujillo et
al. 2009). At first glance this number is pretty similar to the fractions estimated in this Letter (particularly
with those values obtained using the Guo et al. 2013 model). However, the  number estimated by Trujillo et al. (2009) is
uncertain for several reasons. The first one is that the ages of the compact galaxies found in that work are
relatively young ($\sim$ 2 Gyr), this means that they can not be relics of the early Universe. In that sense,
Trujillo et al. results should be understand as an upper limit of the true value (i.e. $<$0.0003; $<1.3\times10^{-7}
\, Mpc^{-3}$). Using a slightly different selection  (M$_\star$$>$8$\times$10$^{10}$\,M$_{\odot}$ and $\Delta$log r$_e$
$<$-0.4 dex with repect to the SDSS stellar mass - size relation from Shen et al. 2003) and focusing only on red
objects (i.e. more likely to be relics from the early Universe), Taylor et al. (2010) found only one dubious
candidate at 0.066$<$z$<$0.12. This is 5000 times less than the expected number density of compact massive galaxies
in the present-day Universe if the high-z massive objects were not evolve neither in size or mass (i.e. 
$\sim$1.5$\times$10$^{-4}$Mpc$^{-3}$ versus the number found  in Taylor et al. 2010,
$<$3$\times$10$^{-8}$Mpc$^{-3}$)

Is it possible that both Trujillo et al. (2009) and Taylor et al. (2010) results can biased against the detection of
massive relics in the present-day Universe? Taylor et al. (2010) have done an extensive study of the potential
biases that affect the SDSS spectroscopic sample and found that their results can not be explained as consequence of
incompleteness. In a recent work, Poggianti et al. (2013)  applying the same selection that in Trujillo et al.
(2009) found 4 old compact massive galaxies in 30.88 deg$^2$  within 0.03$<$z$<$0.11 using the PM2GC (Calvi et al.
2011) sample. This is equivalent to a number density of 1.3$\times$10$^{-5}$Mpc$^{-3}$. This number is significantly
larger than the numbers quoted in Trujillo et al. (2009) and Taylor et al. (2010) studies. Poggianti et al. (2013)
results are based in a sample with a larger spectroscopic completeness. However, the different spectroscopic
completeness of the SDSS survey compared to the PM2GC can not explain a factor of at least 100 in the discrepancy
among these works. Poggianti et al. (2013) indicates that only 25\% of their compact galaxies are not identified
spectroscopically in the SDSS.  It is clear that a more detailed analysis is necessary to solve this enormous
discrepancy.

Although the expected fraction of massive relics galaxies change by a factor of 5 among the different
semi-analytical models, the change in the number density can be as high as a factor of 10. For the discussion and
the comparison with the observations, we take the average values of the different models for the fraction and the
number density of massive relics. The results based on the SDSS survey (Trujillo et al. 2009; Taylor et al. 2010)
found a significant less (more than a factor of 10) number  of relics than what is predicted by the models. If these
observations are confirmed in the future this will be hardly understood within the merging scenario. Due to its
stochastic nature, there should be a number (small but measurable) of relics in the nearby Universe that remains
undetected. Again, if confirmed, the absence of compact massive relic galaxies in the nearby Universe will be a
strong indication that the growth mechanisms of the massive galaxies need something else than merging to produce the
observed massive population or that the number of mergers is much larger than what it is theoretically predicted.
This is odd, as we know that the number of satellites among massive galaxies at all redshift is overpredicted by
those simulations (see e.g. Quilis \& Trujillo 2012). On the contrary, if the number density of massive relics
galaxies is closer to the values quoted by Poggianti et al. (2013) then the number of merging seems to be lower than
what is theoretically expected.  Once more, it is urgent to understand the discrepancy between the observational
results to put a stringent constraint to the question.

\acknowledgments The authors thank Bianca Poggianti for clarifying some aspects of her work and the anonymous referee for useful comments and criticism. This work was  supported by the Spanish Ministerio de
Econom\'{\i}a y Competitividad (MINECO)(grants   AYA2010-21322-C03-01 and AYA2010-21322-C03-02)   and    the   Generalitat   Valenciana   (grant
PROMETEO-2009-103).  The Millennium Simulation databases used in this paper and the web 
application providing online access to them were constructed as part of the activities of the German Astrophysical Virtual Observatory.

\end{document}